\documentclass{osa-article}

\journal{osajournal}

\articletype{Letter}

\usepackage{xcolor}

\begin{document}

\title{Mid- and Far-Infrared Supercontinuum Generation in Bulk Tellurium Spanning from 5.3 $\mu$m to 32 $\mu$m}

\author{Daniel Matteo\authormark{*}, Sergei Tochitsky\authormark{*}, and Chan Joshi}

\address{Department of Electrical Engineering, University of California Los Angeles, California 90095, USA}

\email{\authormark{*}Corresponding authors: danielmatteo@ucla.edu, sergei12@ucla.edu} 



\begin{abstract*}
Supercontinuum generation is performed in the bulk semiconductor tellurium (Te) with a high-power picosecond CO$_2$ laser at peak intensities up to 20 GW/cm$^2$. The spectrum spans from the second harmonic of the pump at 5.3 $\mu$m to 32 $\mu$m. Stimulated Raman scattering along with self-phase modulation and four wave mixing are found to be the main nonlinear optical processes leading to the spectral broadening. Numerical simulations using the experimental conditions indicate that the nonlinear refractive index of Te, $n_{2,\textrm{eff}}$(Te) is about (40 $\pm$ 10) $n_{2,\textrm{eff}}$(GaAs), making this a very promising material for nonlinear optical devices. 
\end{abstract*}

\bigskip

Supercontinuum (SC) generation during the propagation of a short laser pulse through a nonlinear material provides extremely broadband coherent sources spanning multiple octaves in different spectral ranges. Abundant applications of SC sources exist, including spectroscopy (e.g. in the MIR molecular fingerprint region) \cite{vodopyanov:20,zorin:22}, interferometry \cite{kim:02}, pulse compression \cite{nisoli:97}, and seeding of tunable optical parametric amplifiers \cite{cerullo:03}.

Much of the previous work in recent years on SC generation has used fibers as the nonlinear medium. Despite their relatively low Kerr nonlinearity, fibers can provide several advantages for SC generation, including guided propagation over scalable interaction length and the ability to engineer material dispersion profiles, such as in photonic crystal fibers \cite{dudley:06}. Dispersion properties of fibers and a multitude of NIR laser sources together allow for pumping in the anomalous/negative group-velocity dispersion (GVD) regime, where soliton dynamics often play a significant role in both propagation and broadband frequency generation \cite{herrmann:02}.

With more recent development of ultrafast laser sources in the MIR \cite{mirov:15}, there has been a sustained interest in expanding the SC spectral range to longer wavelengths. In particular, specially designed chalcogenide fibers have emerged as a medium for SC generation in the MIR \cite{petersen:14}. Broad transparency windows of these fibers have allowed for spectral broadening to extend as far as 16 $\mu$m covering almost the entire molecular fingerprint region from 2 to 20 $\mu$m \cite{zhao:17}. 

An alternative approach to SC generation uses bulk crystals as the nonlinear medium. While scaling crystal length is typically not practical, optical nonlinearities can be orders of magnitude larger than those in fibers. Historically a predecessor of broadband radiation generation in fibers, bulk SC experiments have been successfully performed in crystals such as GaAs \cite{corkum:85,pigeon:14,lanin:15}, YAG \cite{silva:12}, and ZnSe \cite{werner:19}, which provide naturally wide transparency ranges. So far, bulk SC generation has produced spectral broadening to wavelengths as long as 20 $\mu$m in GaAs \cite{pigeon:14}.

Few materials have natural transparency throughout the long wavelength part of the infrared into the far-infrared. One of the most promising options for expanding SC beyond 20 $\mu$m is the narrow-gap semiconductor Te, transparent from 4-32 $\mu$m. To avoid two-photon absorption, Te should only be pumped at wavelengths longer than 8 $\mu$m, rendering most high-power NIR and MIR laser sources impractical. 

Optically, Te has one of the largest known second order nonlinearity (d = 600 pm/V) of any natural bulk material \cite{mcfee:70}. The Kerr nonlinearity ($n_{2}$) is not known, but band gap scaling \cite{sheik-bahae:91} suggests it can be very large, on the order of 10-100x that of GaAs. This third order nonlinearity is the main force behind SC generation, mediating effects such as self-phase modulation (SPM) four wave mixing (FWM), and stimulated Raman scattering (SRS). Therefore, measurements of SC spectra in Te can allow for an estimate of $n_{2}$. Characterization of Te's nonlinear optical properties are especially relevant due to renewed interest in the material; the unique crystal structure, consisting of helical chains along the c-axis weakly bound to each other in a hexagonal array, and its symmetries have been the topic of several recent studies on exotic magnetic \cite{furukawa:17} and topological properties \cite{sakano:20}. 

In this letter, we report the first experiments on reaching far-infrared wavelengths using supercontinuum generation in a 5mm long Te crystal pumped with a GW peak power picosecond CO$_2$ laser at 10.6 $\mu$m. The generated spectrum spans from the second harmonic at 5.3 $\mu$m to near the edge of the material transparency at 32 $\mu$m. We observe clear spectral features associated with Stokes and anti-Stokes components of SRS. Indeed modeling with the generalized nonlinear Schrodinger equation indicates that SRS along with SPM and FWM are the main nonlinear optical processes that produce the spectral broadening leading to the formation of the supercontinuum as the pump frequency lies in the normal dispersion regime. The modeling also enables an estimate of $n_{2}$ of Te. 


SC generation is performed using a picosecond CO$_2$ pump laser which has been described elsewhere \cite{tochitsky:12}. The central wavelength of the pump is 10.6 $\mu$m, the 10P(20) line of the CO$_2$ molecule gain spectrum. The gain spectrum in high pressure CO$_2$ amplifiers is modulated due to incomplete overlap of adjacent rovibrational lines, leading to generation of a pulse train. This pulse train comprises several 3.5 ps long pulses, separated by 18.5 ps, defined by the $\sim$54 GHz separation of lines in the CO$_2$ molecule. The full pulse train contains up to 30 mJ, but to reduce the duration of the interactions and fluence we focus the beam tightly in air creating an avalanche-ionized plasma shutter. Further details on the plasma shutter operation can be found in \cite{welch:22}. The pulse screened by the air-plasma shutter contains up to 4 mJ, and its temporal structure consists of 3-4 short pulses, as shown in Fig. 1(a). The most energetic pulse contains approximately 50$\%$ of the total energy.

\begin{figure}[h!]
\centering\includegraphics[width=6.5cm]{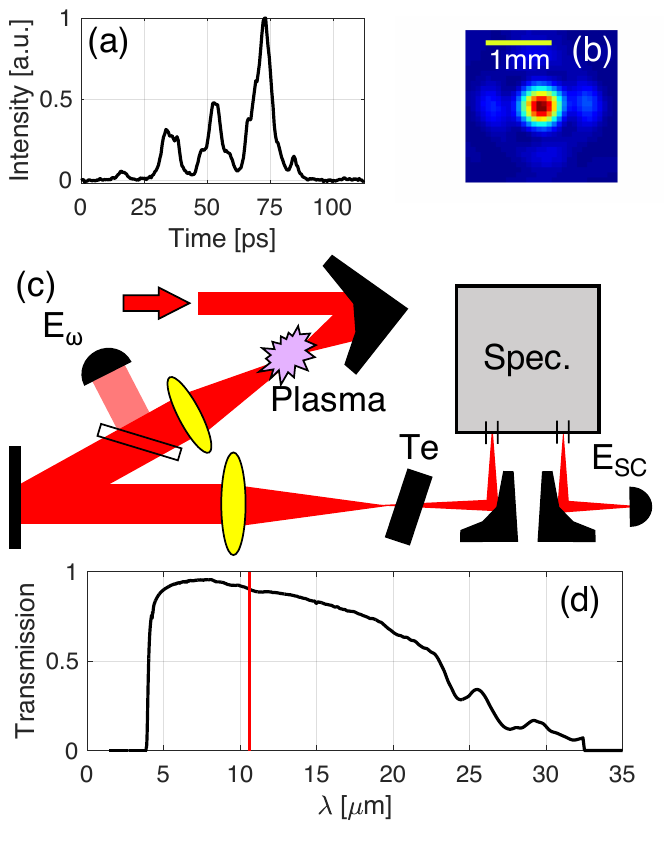}
\caption{(a) Pulse profile used in experiment, measured using a streak camera with an optical upconversion technique. (b) Beam profile on the front surface of the crystal. (c) Experimental setup used to collect and detect SC light. (d) Transmission measured in our 5mm bulk Te crystal, corrected for Fresnel reflection. Pump wavelength is marked. }
\end{figure}

The beam is focused with a 50 cm focal length ZnSe lens to a spot size of $\sim$450 $\mu$m (1/$e^{2}$ radius) on the surface of the Te sample (Fig. 1(b)). Measurement of the beam profile using an IR camera shows that 30$\%$ of the total energy is contained within the beam core. The sample is placed slightly after the beam waist to help mitigate self-focusing. Calculated peak intensity inside the crystal is 20 GW/cm$^2$ when taking into account significant Fresnel reflection. Note that even at such a high fluence of >100 mJ/cm$^2$, no sign of crystal damage was observed over multishot exposure.

The sample is a single Te crystal with dimensions 10x10x5 mm. The transparency of the crystal used in experiments was measured by Fourier transform infrared spectroscopy (Fig. 1(d)). It is cut at a 14$^{\circ}$ angle from the c-axis, allowing for phase matched type I (eeo) second harmonic generation (SHG) of 10.6 $\mu$m light. We pumped the crystal in two different orientations, first with the laser electric field polarized in the nominal phase matching direction. We also rotated the crystal by 90 degrees to detune from the SHG phase matching. 

To measure the SC, we collect the full beam after the crystal and focus it into a spectrometer (Horiba Jobin-Yvon iHR550). Three different diffraction gratings with known wavelength dependent relative efficiency were used to cover the entire Te transparency range. The beam exiting the spectrometer was focused onto a cryogenically cooled 1x1mm HgCdTe detector with known spectral sensitivity peaked at a wavelength of 20 µm. Absolute energy calibration of the detection system was performed using a Gentec calorimeter. 

From the exit of the crystal to the detector, the beam travels through 2.5 m of lab air consisting of approximately 3.5 Torr of H$_2$O, which has strong absorption lines at certain wavelengths beyond 20 $\mu$m \cite{hitran}. Fortunately, there exist several transmission windows between absorption lines making measurements possible in these wavelength windows. For the selected wavelengths, no measureable difference has observed observed with the sealed spectrometer filled with argon or air. 

The experimentally measured SC spectrum in the phase matched orientation is shown in Fig. 2. Each data point is the average of 10-15 shots with calculated intensities binned between 8-20 GW/cm$^2$. The spectrum spans the range from 7.5 to 32 $\mu$m continuously and, considering SHG, extends to 5.3 $\mu$m on the short wavelength side. Gaps in the spectrum around 16.5 and 24 $\mu$m are caused by parasitic ghosts of the diffraction grating making noise associated with the residual pump radiation too high. To the best of our knowledge, this represents the furthest extension of a SC source into the far-infrared spectral region. The total energy contained in the far infrared plateau of the spectrum from 15-32 $\mu$m is measured to be around $\geq$0.5 $\mu$J, or >10$^{-4}$ of the initial pump energy. 

\begin{figure}[h!]
\centering\includegraphics[width=8.5cm]{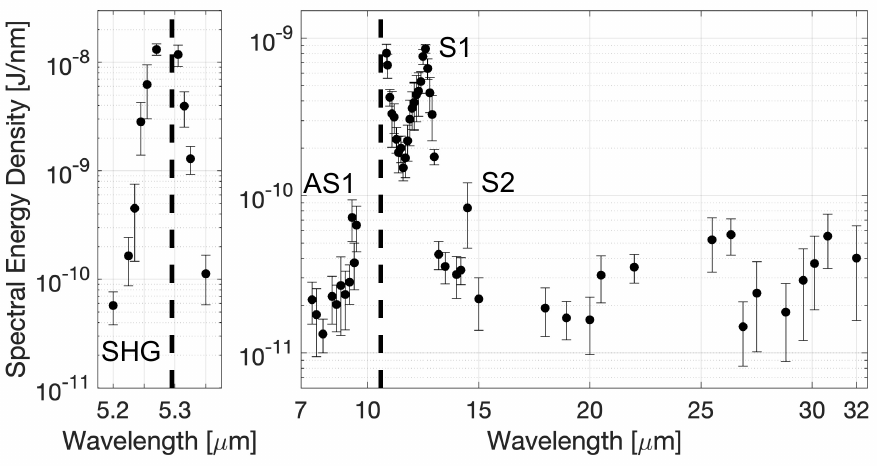}
\caption{Full SC spectrum measured in experiment. Notable features are indicated by dashed lines: SRS Stokes sidebands: S1 and S2, SRS anti-Stokes sideband AS1, and SHG. Pump wavelength and its second harmonic are both marked. Note the different vertical scales for the panels.}
\end{figure}

Stray light was taken into account in the FIR spectral range using a narrow bandpass filter transmitting 85$\%$ of the pump wavelength, and completely absorbing the FIR radiation (Ge substrate). Signal to noise ratio was found to be 2.0 at 22 $\mu$m and 1.4 at 30.7 $\mu$m. 

Notable features are labeled on the figure, including clear SHG, as well as peaks attributed to Stokes (S1, S2) and anti-Stokes (AS1) SRS sidebands. The Raman spectrum of Te has three peaks, but is dominated by the symmetric breathing mode of the Te chain with a frequency of 123 cm$^{-1}$ \cite{qin:20}. Theoretical first and second Stokes sidebands of SRS would appear at 12.2 and 14.3 $\mu$m, which are close to the peaks we observe in experiment. 

According to the literature dispersion data, the zero GVD point of Te exists at 19.7 $\mu$m for E$\parallel$c (e-wave) and 16.6 $\mu$m for E$\perp$c (o-wave) \cite{bhar:76}. Thus, regardless of orientation, the 10.6 $\mu$m pump propagates in the normal, positive GVD regime. Because of this, soliton fission, which often dominates SC spectra generated in the NIR, cannot occur \cite{demircan:07}. 

It is known that modulational instabilities, though typically requiring negative GVD to grow, can experience gain if higher order dispersion parameters are negative \cite{demircan:07}. At 10.6 $\mu$m $\beta_m < 0$ for $m\geq4$, and using a perturbative expansion of the instability gain up to $m=8$ \cite{reeves:03}, gain only exists at wavelengths between 8.5 and 9.6 $\mu$m. Downshifting by these frequencies cannot occur, and no evidence of high-frequency sidebands was found. Thus the observed SC generation is likely caused by SRS in combination with SPM, supported by the relatively strong sideband efficiency.

When the crystal orientation was changed in order to detune the SHG phase matching, the second harmonic component was still present but weaker by an order of magnitude. No significant change in the far infrared SC plateau was found for the two orientations. However, obvious differences were found between the two orientations for the SRS sideband production efficiency. The phase matched orientation was found to generate the 1st Stokes sideband with nearly an order of magnitude more spectral energy density, whereas the non-phase matched orientation produced 1st anti-Stokes and 2nd Stokes SRS sidebands more efficiently. Detailed measurements of the 3 detected SRS sidebands are shown in Fig. 3. 

\begin{figure}[h!]
\centering\includegraphics[width=8.5cm]{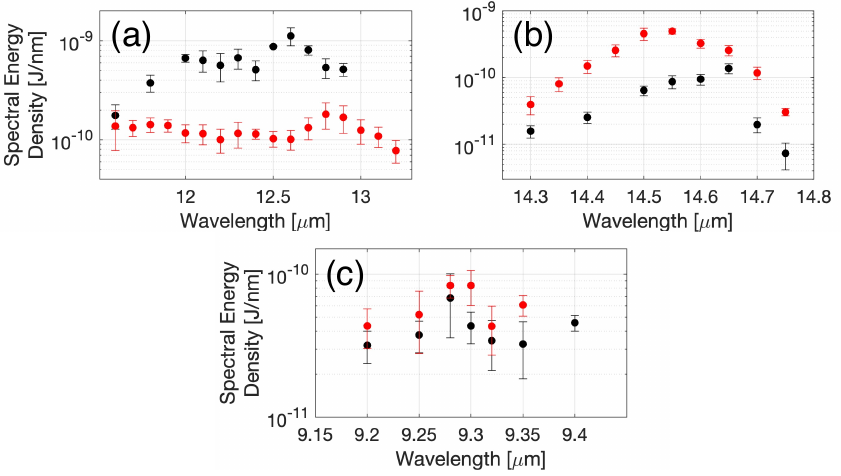}
\caption{Detailed measurements of the (a) 1st Stokes, (b) 2nd Stokes, and (c) 1st anti-Stokes sidebands. Black data is measured in the phase matched orientation, while red data is from the non phase matched orientation.}
\end{figure}

The anisotropy of the Raman response in Te has recently been reported in \cite{jnawali:20}. For E$\parallel$c, the response is weaker compared to E$\perp$c but this alone cannot explain the stark contrast we see in experiment. Possible explanations of this sideband behavior could be over-pumping of the 1st Stokes sideband such that 2nd Stokes and 1st anti-Stokes see enhancement, FWM between the pump and 1st Stokes with an anisotropic $\chi^{(3)}$, or effects of cascaded quadratic nonlinearities. This intricate redistribution of photon energy is yet to be fully understood and requires additional studies using a sample cut such that clean interactions with E$\parallel$c and E$\perp$c are possible.


To better understand the complicated interplay of different $\chi^{(3)}$ phenomena, we performed 1D modeling of SC generation in 5mm of Te by solving the generalized nonlinear Schrodinger equation using the split-step Fourier method \cite{agrawal}. Dispersion is taken from Ref. \cite{bhar:76}, and the absorption spectrum is calculated from the measured transmission data (Fig 1(d)). Third-order nonlinearities are accounted for with a nonlinear response function $R(t) \propto n_{2,\textrm{eff}}\left[(1-f_R)\delta(t) + f_R h_R(t)\right]$,
where $\delta(t)$ represents the instantaneous nonlinear response, $h_R(t)$ gives the delayed Raman response, and $f_R$ is an empirical coefficient describing the strength of the Raman response. Self-steepening is also considered. 

Second-order nonlinear effects and nonlinear absorption are not taken into account; we focused on the long-wavelength SC plateau in the generated SC spectrum. Modeling is performed as a qualitative guide to understand the physical mechanisms behind the SC measured in experiment and estimate $n_{2,\textrm{eff}}$ of Te. 

In the simulations, a train of 3.5 ps long Gaussian pulses approximating Fig. 1(a) is launched through 5mm of Te in the phase matched orientation. The most intense pulse is set at 20 GW/cm$^2$, and noise is modeled in the time domain with magnitude proportional to the pulse intensity in each discretization bin \cite{dudley:06}. The Raman resopnse $h_R(t)$ was constructed by fitting a response function \cite{agrawal} to the known Raman spectrum \cite{qin:20} in the frequency domain. 
 
The only free parameters in the model are $n_{2,\textrm{eff}}$ and $f_R$. Therefore by comparing the spectral broadening and spectral energy density of the numerical spectrum to the measured one, an estimate of $n_{2,\textrm{eff}}$ can be made. 

\begin{figure}[h!]
\centering\includegraphics[width=8.5cm]{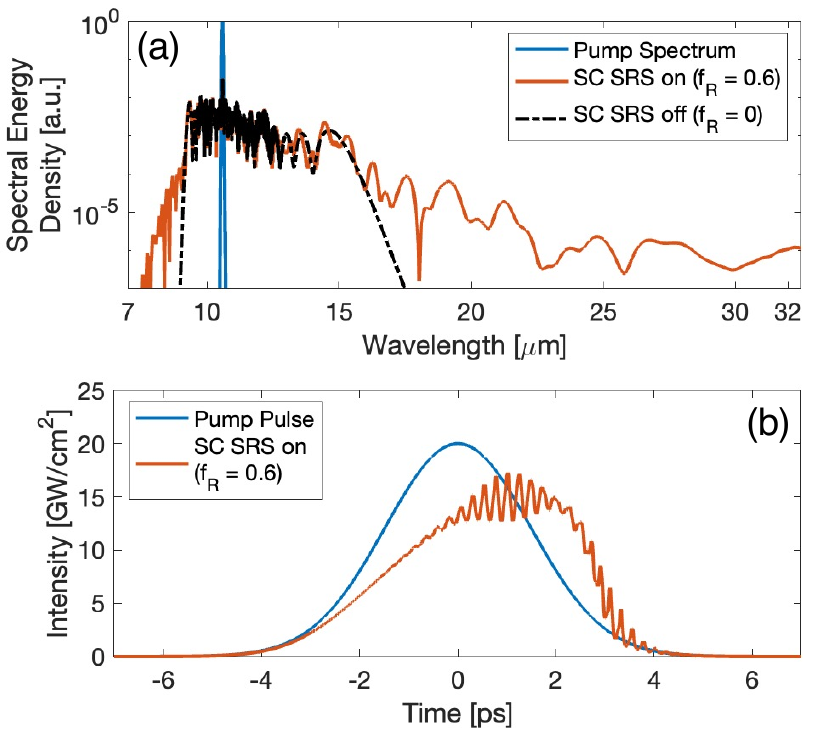}
\caption{(a) Numerical spectra calculated in Te for $n_{2,\textrm{eff}}$ = 40x$n_{2,\textrm{eff}}$(GaAs). SRS is shown to be the principle driver of the SC generation. Smoothing of the spectra is applied only to reduce the modulation depth near the pump wavelength; it has no other effect on the spectrum. (b) Corresponding temporal profile of the spectra in (a). Only the most intense pulse in the pulse train is shown.}
\end{figure}

The simulated spectrum at the exit face of the sample is shown in Fig. 4(a). We found that the observed and simulated spectra extended out to similar wavelengths and qualitatively had similar shape of the long wavelength plateau when  $n_{2,\textrm{eff}}$ = 40 $n_{2,\textrm{eff}}$(GaAs) and $f_R$ = 0.6. Fig. 4(a) also shows a simulation with identical parameters, but with the Raman response turned off ($f_R$ = 0). In that case, only FWM, SPM, and self-steepening contribute to spectral broadening, and the spectral extent is significantly less than what we observe in experiment. This further supports the hypothesis that SRS plays an important role in extending the SC range in Te. 
 
Fig 4(b) shows the time domain corresponding to the spectra in Fig. 1(a) (only the most intense pulse is shown). Compared to the pump pulse shape, the pulse after propagation has steepened on the trailing edge. Modulation is visible due to strong SRS, but full pulse splitting has yet to occur. Thus modeling indicates that despite very high nonlinearity a 5 mm Te crystal can produce a coherent mid-and far-IR beam due to lack of pulse splitting \cite{alfano:06}.

By tuning $n_{2,\textrm{eff}}$ in the simulations we can establish approximate numerical uncertainty bounds based on relative conversion efficiency of the SC in the far infrared. We find $n_{2,\textrm{eff}}$(Te) = $(40 \pm 10)n_{2,\textrm{eff}}$(GaAs). Considering $n_{2,\textrm{eff}}$ of GaAs to be 4$\pm 2\times 10^{-14}$ cm$^{2}$/W at 10.6 $\mu$m \cite{wynne:69}, the nonresonant nonlinearity of 1.6$\times 10^{-12}$ in Te is extraordinarily large, rivaling the largest known in any natural material. 

It should be noted that the contribution of bound electron and free carrier nonlinearities to the effective nonlinearity remains a point of interest and requires additional studies. Intrinsic Te is p-type \cite{caldwell:59}, so free carrier nonlinearities likely play a significant role in the overall nonlinear response. 

In conclusion, we have demonstrated supercontinuum generation from 5-32 $\mu$m in bulk Te pumped by high intensity 10 $\mu$m laser pulses. It is mainly driven by stimulated Raman scattering. An ultrabroadband source such as this could be useful for molecular spectroscopy. In addition, by matching the measured spectral span with numerical data we estimate $n_{2,\textrm{eff}}$ in Te to be $\sim$40x that of GaAs, making it a promising material for future infrared nonlinear photonic devices.
 
\begin{backmatter}
\bmsection{Funding}
Office of Naval Research Multidisciplinary University Research Initiative (N00014-17-1-2705).

\bmsection{Disclosures} The authors declare no conflicts of interest.

\bmsection{Data availability} Data underlying the results presented in this paper are not publicly available at this time but may be obtained from the authors upon reasonable request.

\end{backmatter}


\bibliography{references}

\begin{thebibliography}{10}
\newcommand{\enquote}[1]{``#1''}

\bibitem{vodopyanov:20}
K.~L. Vodopyanov, \emph{Laser-based mid-infrared sources and applications}
  (John Wiley $\&$ Sons, Hoboken, NJ, 2020).

\bibitem{zorin:22}
I.~Zorin, P.~Gattinger, A.~Ebner, and M.~Brandstetter, \enquote{{Advances in
  mid-infrared spectroscopy enabled by supercotninuum laser sources},}
  {\protect\JournalTitle{Optics Express}} \textbf{30}, 5222--5254 (2022).

\bibitem{kim:02}
K.~Y. Kim, I.~Alexeev, and H.~M. Milchberg, \enquote{{Single-shot
  supercontinuum spectral interferometry},} {\protect\JournalTitle{Applied
  Physics Letters}} \textbf{81}, 4124 (2002).

\bibitem{nisoli:97}
M.~Nisoli, S.~D. Silvestri, O.~Svelto, R.~Szipocs, K.~Ferencz, C.~Spielmann,
  S.~Sarania, and F., \enquote{{Compression of high-energy laser pulses below 5
  fs},} {\protect\JournalTitle{Optics Letters}} \textbf{22}, 522--524 (1997).

\bibitem{cerullo:03}
G.~Cerullo and S.~D. Silvestri, \enquote{{Ultrafast optical parametric
  amplifiers},} {\protect\JournalTitle{Rev. Sci. Instr.}} \textbf{74}, 1--18
  (2003).

\bibitem{dudley:06}
J.~M. Dudley, G.~Genty, and S.~Coen, \enquote{{Supercontinuum generation in
  photonic crystal fiber},} {\protect\JournalTitle{Rev. Mod. Phys.}}
  \textbf{78}, 1135--1184 (2006).

\bibitem{herrmann:02}
J.~Herrmann, U.~Griebner, N.~Zhavoronkov, A.~Husakou, D.~Nickel, J.~C. Knight,
  W.~J. Wadsworth, P.~S.~J. Russel, and G.~Korn, \enquote{{Experimental
  evidence for supercontinuum generation by fission of higher-order solitons
  photonic fibers},} {\protect\JournalTitle{Phys. Rev. Lett.}} \textbf{88},
  173901 (2002).

\bibitem{mirov:15}
S.~B. Mirov, V.~V. Fedorov, D.~Martyshkin, I.~S. Moskalev, M.~Mirov, and
  S.~Vasilyev, \enquote{{Progress in mid-IR lasers based on Cr and Fe-doped
  II-VI chalcogenides},} {\protect\JournalTitle{IEEE. J. Sel. Top. Quantum
  Electron.}} \textbf{21}, 292 (2015).

\bibitem{petersen:14}
C.~R. Petersen, U.~Moller, I.~Kubat, B.~Zhou, S.~Dupont, J.~Ramsay, T.~Benson,
  S.~Sujecki, N.~Abdel-Moneim, Z.~Tang, D.~Furniss, A.~Seddon, and O.~Bang,
  \enquote{{Mid-infrared supercontinuum covering the 1.4-13.3 $\mu$m molecular
  fingerprint region using ultra-high NA chalcogenide step-index fibre},}
  {\protect\JournalTitle{Nature Photonics}} \textbf{8}, 830--834 (2014).

\bibitem{zhao:17}
Z.~Zhao, B.~Wu, X.~Wang, Z.~Pan, Z.~Liu, P.~Zhang, X.~Shen, Q.~Nie, S.~Dai, and
  R.~Wang, \enquote{{Mid-infrared supercontinuum covering 2.0-16 $\mu$m in a
  low-loss telluride single-mode fiber},} {\protect\JournalTitle{Laser
  Photonics Reviews}} \textbf{11}, 1700005 (2017).

\bibitem{corkum:85}
P.~B. Corkum, P.~P. Ho, R.~R. Alfano, and J.~T. Manassah, \enquote{{Generation
  of infrared supercontinuum covering 3-14 $\mu$m in dielectrics and
  semiconductors},} {\protect\JournalTitle{Optics Letters}} \textbf{10},
  624--626 (1985).

\bibitem{pigeon:14}
J.~J. Pigeon, S.~Y. Tochitsky, C.~Gong, and C.~Joshi, \enquote{{Supercontinuum
  generation from 2 to 20 $\mu$m in GaAs pumped by picosecond CO$_2$ laser
  pulses},} {\protect\JournalTitle{Optics Letters}} \textbf{39}, 3246--3249
  (2014).

\bibitem{lanin:15}
A.~A. Lanin, A.~A. Voronin, E.~A. Stepanov, A.~B. Fedotov, and A.~M. Zheltikov,
  \enquote{{Multioctave, 3--18 $\mu$m sub-two-cycle supercontinua from
  self-compressing, self-focussing soliton transients in a solid},}
  {\protect\JournalTitle{Optics Letters}} \textbf{40}, 974--977 (2015).

\bibitem{silva:12}
F.~Silva, D.~R. Austin, A.~Thai, M.~Baudisch, M.~Hemmer, D.~Faccio,
  A.~Couairon, and J.~Biegert, \enquote{{Multi-octave supercontinuum generation
  from mid-infrared filamentation in a bulk crystal},}
  {\protect\JournalTitle{Nature Communications}} \textbf{3}, 807 (2012).

\bibitem{werner:19}
K.~Werner, M.~G. Hastings, A.~Schweinsberg, B.~L. Wilmer, D.~Austin, C.~M.
  Wolfe, M.~Kolesik, T.~R. Ensley, L.~Vanderhoef, A.~Valenzuela, and
  E.~Chowdhury, \enquote{{Ultrafast mid-infrared high harmonic and
  supercontinuum generation with $n_{2}$ characterization in zinc selenide},}
  {\protect\JournalTitle{Optics Express}} \textbf{27}, 2867--2885 (2019).

\bibitem{mcfee:70}
J.~H. McFee and G.~D. Boyd, \enquote{{Redetermination of the of the nonlinear
  optical coefficients of Te and GaAs by comparson wth Ag$_3$SbS$_3$},}
  {\protect\JournalTitle{Appl. Phys. Lett.}} \textbf{17}, 57--59 (1970).

\bibitem{sheik-bahae:91}
M.~Sheik-Bahae, D.~C. Hutchigns, D.~J. Hagan, and E.~W.~V. Stryland,
  \enquote{{Dispersion of bound electronic nonlinear refraction in solids},}
  {\protect\JournalTitle{IEEE J. Quantum Electron}} \textbf{27}, 1296 (1991).

\bibitem{furukawa:17}
T.~Furukawa, Y.~Shimokawa, K.~Kobayashi, and T.~Itou, \enquote{{Observation of
  current-induced bulk magnetization in elemental telllurium},}
  {\protect\JournalTitle{Nature Communications}} \textbf{8}, 954 (2017).

\bibitem{sakano:20}
M.~Sakano and M.~Hirayama, \enquote{{Radial spin texture in elemental tellurium
  with chiral crystal structure},} {\protect\JournalTitle{PRL}} \textbf{124},
  136404 (2020).

\bibitem{tochitsky:12}
S.~Y. Tochitsky, J.~J. Pigeon, D.~J. Haberberger, C.~Gong, and C.~Joshi,
  \enquote{{Amplification of multi-gigawatt 3 ps pulses in an atmospheric
  CO$_2$ laser using ac Stark effect},} {\protect\JournalTitle{Opt. Express}}
  \textbf{20}, 13762--13768 (2012).

\bibitem{welch:22}
E.~Welch, D.~Matteo, S.~Tochitsky, G.~Louwrens, and C.~Joshi,
  \enquote{{Observation of breakdown wave mechanism in avalanche ionization
  produced atmospheric plasma generated by a picosecond CO$_2$ laser},}
  {\protect\JournalTitle{Physics of Plasmas}} \textbf{29}, 053504 (2022).

\bibitem{hitran}
I.~E. Gordon and et~al., \enquote{{The HITRAN2020 molecular spectroscopic
  database},} {\protect\JournalTitle{J. Quant. Spectrosc. Radiat. Transf.}}
  \textbf{277}, 107949 (2022).

\bibitem{qin:20}
J.-K. Qin, P.-Y. Liao, M.~Si, S.~Gao, G.~Qui, J.~Jian, Q.~Wang, S.-Q. Zhang,
  S.~Huang, A.~Charnas, Y.~Wang, M.~J. Kim, W.~Wu, X.~Xu, H.-Y. Wang, L.~Yang,
  Y.~K. Yap, and P.~D. Ye, \enquote{{Raman response and transport properties of
  tellurium atomic chains encapsulated in nanotubes},}
  {\protect\JournalTitle{Nature Electronics}} \textbf{3}, 141--147 (2020).

\bibitem{bhar:76}
G.~C. Bhar, \enquote{{Refractive index interpolation in phase-matching},}
  {\protect\JournalTitle{Applied Optics}} \textbf{15}, 305--307 (1976).

\bibitem{demircan:07}
A.~Demircan and U.~Bandelow, \enquote{{Analysis of the interplay between
  soliton fission and modulation instability in supercontinuum generation},}
  {\protect\JournalTitle{Appl. Phys. B}} \textbf{86}, 31--39 (2007).

\bibitem{reeves:03}
W.~H. Reeves, D.~V. Skryabin, F.~Biancalana, J.~C. Knight, P.~S.~J. Russel,
  F.~G. Omenetto, A.~Efimov, and A.~J. Taylor, \enquote{{Transformation and
  control of ultrashort pulses in dispersion-engineered},}
  {\protect\JournalTitle{Nature}} \textbf{424}, 511--515 (2003).

\bibitem{jnawali:20}
G.~Jnawali, \enquote{{Ultrafast photoinduced band splitting and carrier
  dynamics in chiral tellurium nanosheets},} {\protect\JournalTitle{Nature
  Communications}} \textbf{11}, 3991 (2020).

\bibitem{agrawal}
G.~P. Agrawal, \emph{{Nonlinear fiber optics}} (Academic Press, 2007), 4th ed.

\bibitem{alfano:06}
R.~R. Alfano, ed., \emph{The Supercontinuum Laser Source} (Springer, Berlin,
  2006).

\bibitem{wynne:69}
J.~J. Wynne, \enquote{{Optical third-order mixing in GaAs, Ge, Si, and InAs},}
  {\protect\JournalTitle{Phys. Rev.}} \textbf{178}, 1295--1303 (1969).

\bibitem{caldwell:59}
R.~S. Caldwell and H.~Y. Fan, \enquote{Optical properties of tellurium and
  selenium,} {\protect\JournalTitle{Phys. Rev.}} \textbf{114}, 664--675 (1959).

\end{thebibliography}

\end{document}